\documentclass[useAMS,usenatbib]{mn2e}
\usepackage{amssymb,amsmath}
\usepackage{url}
\usepackage{xcolor}
\usepackage{epstopdf}

\usepackage{graphicx}
\usepackage{bmpsize}
\DeclareGraphicsExtensions{.pdf,.png,.jpg}

\usepackage{color}

\title[Interstellar dust in astrospheres]{Non-monotonic spatial distribution of the interstellar dust in astrospheres: finite gyroradius effect}
\author[O. A.~Katushkina et al.]{O. A.~Katushkina$^{1}$\thanks{E-mail: okat@iki.rssi.ru},
 D. B.~Alexashov$^{1,2}$, V. V.~Izmodenov$^{1,2,3}$
 \newauthor 
 and  V. V.~Gvaramadze$^{1,4,5}$ \\
$^{1}$Space Research Institute of Russian Academy of Sciences, Profsoyuznaya Str. 84/32, Moscow, 117335, Russia\\
$^{2}$Institute for Problems in Mechanics, prosp. Vernadskogo 101, block 1, Moscow, 119526, Russia\\
$^{3}$Lomonosov Moscow State University, GSP-1, Leninskie Gory, Moscow, 119991, Russia\\
$^{4}$Sternberg Astronomical Institute, Moscow State University, Universitetskij Pr. 13, Moscow 119992, Russia\\
$^{5}$Isaac Newton Institute of Chile, Moscow Branch, Universitetskij Pr. 13, Moscow 119992, Russia \\
}
\begin{document}

\def\vecv{\mbox{$\textbf{v}_p$}}
\def\vecV{\mbox{$\textbf{V}$}}
\def\vecB{\mbox{$\textbf{B}$}}
\def\vecq{\mbox{$\textbf{q}$}}
\def\vecQ{\mbox{$\textbf{Q}$}}

\date{Accepted 2016 November 1. Received 2016 October 7; in original form 2016 September 6}

\pagerange{1573 -- 1585} \pubyear{2017}

\maketitle

\label{firstpage}

\begin{abstract}
High-resolution mid-infrared observations of astrospheres
show that many of them have filamentary (cirrus-like)
structure. Using numerical models of dust dynamics in
astrospheres, we suggest that their filamentary structure might be
related to specific spatial distribution of the interstellar dust around
the stars, caused by a gyrorotation of charged dust grains in the
interstellar magnetic filed. Our numerical model describes the dust dynamics in astrospheres under an influence
of the Lorentz force and assumption of a constant dust charge. Calculations are performed
for the dust grains with different sizes separately.
It is shown that non-monotonic spatial dust distribution
(viewed as filaments) appears for dust grains with the period of
gyromotion comparable with the characteristic time scale of the
dust motion in the astrosphere. Numerical modelling demonstrates that number of filaments depends on
charge-to-mass ratio of dust.


\end{abstract}


\section{Introduction}
\label{sec:int}


Astrospheres are the structures formed around stars because of
interaction between the stellar wind (SW) and the surrounding
interstellar medium (ISM). One well-known example of an
astrosphere is the heliosphere around the Sun. The shape of a
particular astrosphere is determined by the properties of the SW
and the local ISM. Differences in these properties may lead to a
wide range of shapes of observed astrospheres (Sahai \&
Chronopoulos 2010; Gvaramadze et al. 2011a; Peri et al. 2012;
Decin et al. 2012; Cox et al. 2012; Peri, Benaglia \& Isequilla
2015; Kobulnicky et al. 2016).

\begin{figure}
\includegraphics[width=10cm]{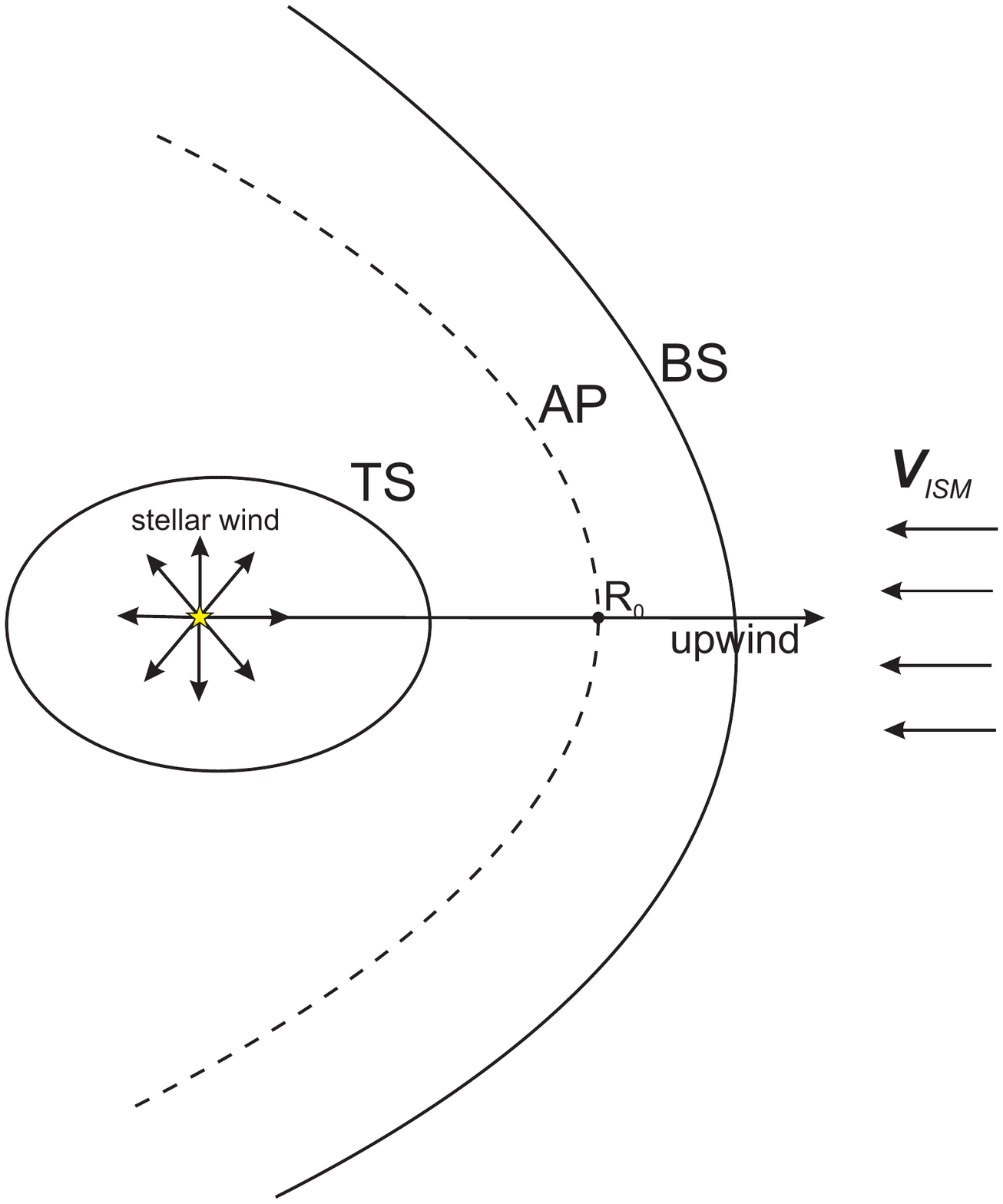}
\caption{Schematic structure of an astrosphere. TS, AP and BS are,
respectively, the termination shock, the astropause and the bow shock.
The upwind is the direction of star's motion relative to the surrounded ISM.
$R_0$ is the stand-off distance.}
\label{interface}
\end{figure}

The region of interaction between the supersonic SW and the
supersonic ISM flow contains three discontinuities (see
Fig.~\ref{interface}): (1) the stellar wind termination shock (TS), separating the
region of freely flowing wind from the shocked wind, (2) the
tangential discontinuity or the astropause (AP), which separates
the material of the SW from that of the ISM, and at which the SW
ram pressure is balanced by the ram pressure of the ISM, and (3)
the forward shock or the bow shock (BS), separating the shocked
ISM from the unperturbed one. For the first time, such a two-shock
structure was considered for the heliosphere by Baranov,
Krasnobaev \& Kulikovskii (1970), who numerically obtained the
shape of the AP under the thin-layer approximation.
Later on, Wilkin (1996) applied this theory for the shape
of the stellar wind BSs using the same approximation.

The thin-layer solution provides a good description of the shape
of the BSs with efficient cooling, for which the thickness
of the interaction region between the stellar wind and the ISM is
much smaller than the characteristic scale of the BS.
In this case the distance from the star to the BS at nose part of astrosphere is almost equal to the
stand-off distance, $R_0$ that is the distance
from the star to the AP in the upwind direction (Fig.~\ref{interface}).
 For a wind-blowing
star moving with velocity $\textbf{v}_*$ relative to the ISM of mass
density $\rho_{\rm ISM}$, the stand-off distance is given by
\begin{equation}
\label{R0} R_0=\sqrt{\frac{\dot{M} v_{\rm w}}{4\pi \rho_{\rm ISM}
v_* ^2}} \, ,
\end{equation}
where $\dot{M}$ and $v_{\rm w}$ are the stellar wind mass-loss
rate and velocity, respectively.


The assumption that the BS forms at the distance $R_0$ from the
star is widely exploited in the astrophysical literature (e.g. to
determine the properties of the stellar wind or the local ISM;
Kobulnicky, Gilbert \& Kiminki 2010; Gvaramadze, Langer \& Mackey
2012; Peri et al. 2012) and the entire interaction region between
the SW and the ISM is often called as the bow shock. In reality, however,
the thickness of this region could be comparable with the
characteristic scale of the astrosphere (e.g. van Buren 1993; Comer\'on \&
Kaper 1998). To avoid the terminological confusion, we call the
two-shock structure as astrosphere and use the term ``bow shock"
in its classical gasdynamic and magnetohydrodynamic (MHD) meaning.



The astrospheres can be observed at various wavelengths, of which
the mid-infrared waveband is most appropriate for their detection
(van Buren et al. 1995; Gvaramadze \& Bomans 2008; Peri et al.
2012). Numerical modelling of astrospheres around OB stars show
that their infrared emission is mostly due to re-radiation of the
starlight by the circumstellar and interstellar dust grains
accumulated in the astrospheres (Meyer et al. 2014). Thus, to
interpret the available images of astrospheres and to understand
their gasdynamic or MHD structure one needs to analyze the dust
distribution within them.

\begin{figure*}
\includegraphics[width=16cm]{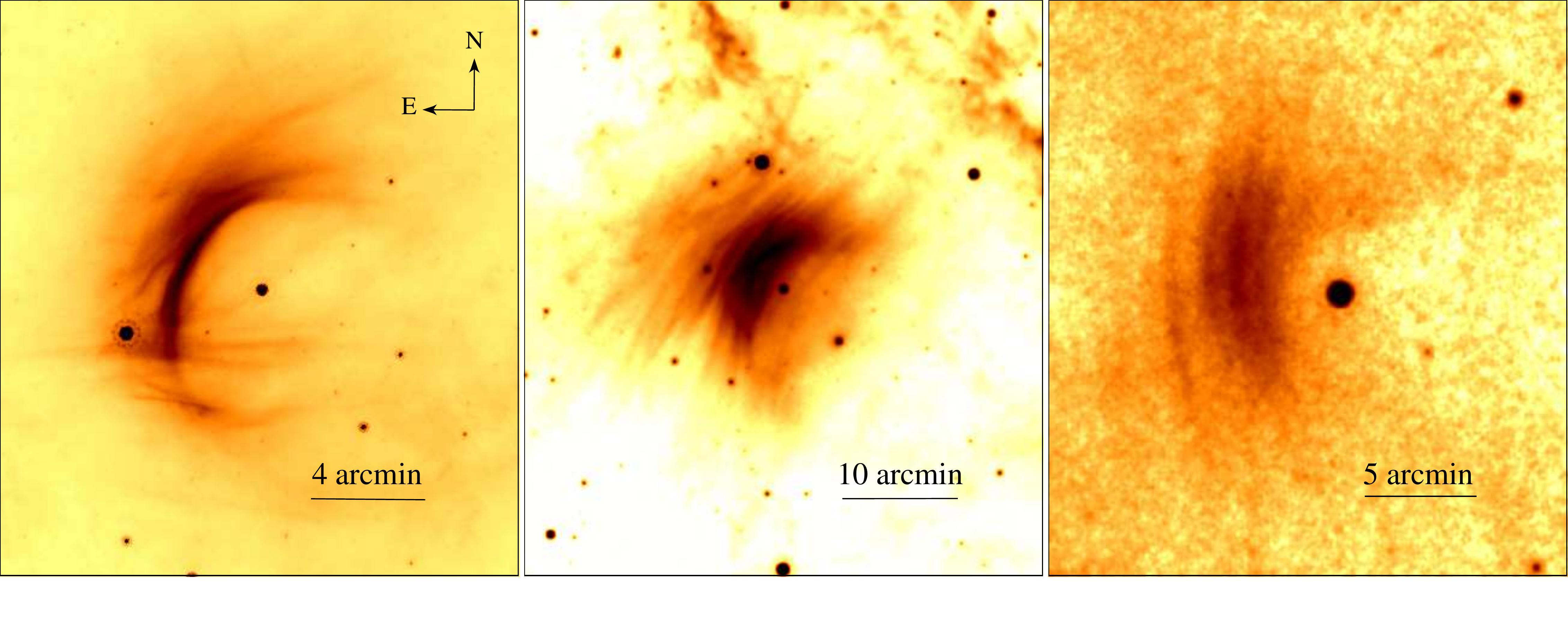}
\label{3stars}
\caption{{\it Spitzer} 24\,$\mu$m (left-hand panel) and {\it WISE} 22\,$\mu$m (middle
and right-hand panels) images of astrospheres associated with three early B stars,
respectively, $\kappa$\,Cas, $\theta$\,Car and $\beta$\,CMa.
The orientation of the images is the same.}
\label{3stars}
\end{figure*}

The high angular resolution of recent infrared surveys carried
out with the {\it Spitzer Space Telescope}, {\it Wide-field
Infrared Survey Explorer} ({\it WISE}) and {\it Herschel Space
Observatory} allowed to resolve the fine structure of numerous
astrospheres and showed that many of them have filamentary
(cirrus-like) structure (France, McCandliss \& Lupu 2007;
Gvaramadze et al. 2011a,b; Peri et al. 2012; Decin et al. 2012; see
also Fig.~\ref{3stars}).
The origin of this structure is not well understood. One can
speculate that it is due to interstellar dust grains aligned with
the local interstellar magnetic field, which re-radiate the light
of central stars of the astrospheres (Gvaramadze et al. 2011b) or
due to time dependent variations of the stellar wind parameters
(Decin et al. 2012), or that it is caused by some kind of
instabilities in the astrospheres (Dgani, Van Buren \&
Noriega-Crespo 1996).

In this paper, we present a new possible physical mechanism
for formation of the cirrus-like structure of some astrospheres.
Our results based on the three dimensional (3D)
modelling of interstellar dust distribution in astrospheres
produced by stars moving through the ISM with a large-scale
homogeneous magnetic field. We show that the dust distribution
strongly depends on the charge-to-mass ratio of dust grains as
well as on the orientation of the magnetic field with respect to
the direction of stellar motion. We also show that synthetic dust column density maps have
filamentary structure
 if the dust gyroperiod
 is comparable with the characteristic time scale of the
dust motion in the astrosphere. This could be an
explanation of the observational data.

The paper is organized as follows. In Section\,\ref{sec:for}, we
describe our numerical 3D stationary MHD model of an astrosphere
and the dust distribution within and around it. Results of the
3D simulations are presented in Section\,\ref{sec:res}.
Section\,\ref{sec:discussion} discusses limitations of our model.
Summary and conclusions are presented in Section\,\ref{sec:sum}.

\section{Mathematical formulation of the problem}
\label{sec:for}

\subsection{Numerical model for the plasma distribution in the astrosphere}
\label{sec:num}


At first, we present a 3D MHD model of the astrosphere around a
wind-blowing star moving with velocity $\textbf{v}_*$ through the uniformly
magnetized ISM. Both the stellar wind and the ISM are assumed to
be fully ionized. For the sake of simplicity, we restrict
ourselves to the framework of ideal MHD and will not consider
radiative cooling and thermal conduction, although these processes
might be important for some astrospheres \citep[see,
e.g.,][]{falle_1975, weaver_etal_1977, meyer_etal_2014}. The
governing equations are the steady-state ideal MHD equations:
\begin{eqnarray}
\label{mag1} &&
\nabla\cdot(\rho \vecv)=0,
\nonumber \\
&&  \nabla\cdot\left[\rho
\vecv\vecv+\left(p+\frac{B^2}{8\pi}\right)\/\/\/\/\mathsf{I}
-\frac{\vecB\vecB}{4\pi}\right] = 0,  \nonumber \\
&& \nabla\cdot\left[\left(\epsilon+p+\frac{B^2}{8\pi}\right)\vecv
-\frac{(\vecv\cdot\vecB)}{4\pi}\vecB\right] = 0
, \\
&& \nabla\cdot\left[\vecv\vecB-\vecB\vecv
\right]=0, \nonumber \\
&& \nabla\cdot \vecB = 0, \nonumber
\end{eqnarray}
where the variables have their usual meanings as $\rho$ (plasma mass
density), $\vecv$ (plasma velocity), $p$ (thermal pressure), $\vecB$
(magnetic field), $\epsilon=\rho v^2/2 +p/(\gamma-1)$ (kinetic
plus internal energy per unit volume), $\gamma$ (adiabatic index),
and $\mathsf{I}$ is the unity tensor.

The inner boundary conditions for the uniform, spherically
symmetric stellar wind taken at a sphere of radius $r_s$ (inside TS) are the
following: the stellar wind velocity is $v_{\rm w}$, the density
is $\rho_{\rm w}=\dot{M}/(4\pi v_{\rm w} r_s^2$), and the thermal
pressure is $p_{\rm w}$. The stellar wind is assumed to be
non-magnetized. The unperturbed ISM is assumed to be uniformly
magnetized with constant density ($\rho_{\rm ISM}=m_{\rm p}n_{\rm ISM}$, where $m_{\rm p}$ is a mass of proton,
$n_{\rm ISM}$ is ISM number density), pressure
($p_{\rm ISM}$) or temperature ($T_{\rm ISM}$), and magnetic field
($\vecB_{\rm ISM}$). In the frame of reference of the star the ISM flows
towards the star with velocity of $\vecV_{\rm ISM}=-\textbf{v}_*$.

It can be shown that dimensionless solution of the formulated
problem depends only on the sonic Mach number in the
stellar wind, $M_{\rm w}=v_{\rm w}/\sqrt{\gamma p_{\rm
w}/\rho_{\rm w}}$, the sonic and Alfvenic Mach numbers in the ISM,
respectively, $M_{\rm ISM}=V_{\rm ISM}/\sqrt{\gamma p_{\rm
ISM}/\rho_{\rm ISM}}$ and $M_{\rm A,ISM}=V_{\rm ISM}\sqrt{4\pi\rho_{\rm ISM}}/B_{\rm ISM}$, and the angle,
$\alpha$, between $\vecV_{\rm ISM}$ and $\vecB_{\rm ISM}$. In this
work, we consider only hypersonic ($M_{\rm w}\gg1$) winds,
 for which the structure of their astrospheres does not
depend on the exact value of $M_{\rm w}$.

%
%
%
More details of the numerical method used to model the stellar
wind-ISM interaction region can be found in
\citet{izmod_alexash_2015}.

\subsection{Model of the dust distribution in astrospheres}
\label{sec:dust}

To calculate the dust distribution in the astrosphere we use a kinetic model developed previously for dust in the heliosphere and described in detail
by \citet{alexash_etal_2016}.
In this work we consider the interstellar dust outside the astropause, where the filaments can be formed.
A brief description of the model is following.


The dynamics of the interstellar charged dust grains outside of
the astropause is determined by the Lorentz force $\textbf{F}_{\rm
L}=q[(\textbf{v}_d-\textbf{v}_p)\times \textbf{B}]$, where $q$
is the charge of a dust grain,
$\textbf{v}_d$ is the local velocity of dust, and $\textbf{v}_p$
is the local velocity of plasma. Since we assume that the stellar
wind is non-magnetized, $F_{\rm L}=0$ inside the astropause. For
massive hot stars the forces of gravitational attraction and
radiation repulsion could be important even far away from the
star. For example, \citet{ochsendorf_etal_2014} have shown that
the stellar radiation pressure of weak-wind B stars can sweep away
the interstellar dust and force it to flow around the moving
stars. In this study, to avoid mixing of different physical
effects, we consider only the
electromagnetic forces and neglect the radiation pressure and
gravitation force.

Thus, trajectory of a dust grain is described by the following
equation (in CGS units):
\begin{equation}
\frac{d \textbf{v}_d}{dt} =
\frac{q}{m \cdot c_0}[(\textbf{v}_d-\textbf{v}_p)\times \textbf{B}],
\label{motion}
\end{equation}
were, $m$ is the mass of a dust grain, $c_0$ is the speed of light, $\textbf{v}_d=\dot{\textbf{x}}_d$, $\textbf{x}_d$ is the
position of a dust grain, and the plasma parameters $\textbf{v}_p$
and $\textbf{B}$ are taken from the MHD model of the astrosphere
described in Section\,\ref{sec:num}.

Initial conditions for equation~(\ref{motion}) are defined in the
unperturbed ISM, where we assume $\textbf{v}_d =\textbf{v}_p =
\vecV_{ISM}$ (i.e. dust moves together with plasma), and, therefore, the Lorentz force is equal to zero.

In dimensionless form the equation of motion (\ref{motion})
becomes:
\begin{equation}
\label{motion_br} \dot{\hat{\textbf{v}}}_d = a_{d}
[(\hat{\textbf{v}}_d-\hat{\textbf{v}}_p)\times \hat{\textbf{B}}],
\end{equation}
where the hatted variables are nondimensional and where there is
only one parameter:
\begin{equation}
\label{ad}
 a_{d}=\frac{q}{m} \cdot \frac{ R_0 \cdot \sqrt{\rho_{ISM}}}{c_0}.
\end{equation}
This is the dimensionless charge-to-mass dust ratio. Here, the charge $q$ is expressed in CGS-units ($[q]=g^{1/2}\,cm^{3/2}\,s^{-1}$), $R_0$ is character distance of the astrosphere that is the stand-off distance
defined in the Introduction.



We do not consider processes of dust charging and destruction,
i.e. the charge and the mass of the dust grains are assumed to be
constant. Correspondingly, the parameter $a_d$ is constant for
each dust grain and does not change along the trajectory. Note
that $a_d\sim r_d^{-2}$ (because $q \sim r_d$, and $m \sim r_d^3$), where $r_d$ is the radius of a dust grain,
i.e. $a_d$ is larger for smaller grains and vice versa.


To calculate the dust distribution the following procedure is used.
We split 3D space by computational cells (plasma parameters and magnetic field are known and assumed
to be constant for each grid cell), run a set of dust grains trajectories and then
calculate dust number density ($n_d$) and averaged velocity ($\textbf{V}_{d,av}$) in each cell by counting the individual particles
and their velocities and averaging over the cell.


In
the next Section we present the results of modelling of the dust
number density obtained for different values of parameter $a_d$.

\section{Results}
\label{sec:res}

In this work, we consider the effect of interstellar magnetic
field on the dust distribution in astrospheres for two limiting
cases of maximum and minimum possible thickness of the layer
between the astropause and the bow shock. In the first so called
``perpendicular case'' the maximum thickness of the layer is
achieved for adiabatic ($\gamma=5/3$) flows with $\vecB_{\rm ISM}
\perp \vecV_{\rm ISM}$ and $M_{\rm ISM}<M_{\rm A,ISM}$. In the
second -- ``parallel case'' the thickness of the layer is minimum
for isothermal ($\gamma\approx1$) flows with  $\vecB_{\rm ISM}
\parallel \vecV_{ISM}$. For this orientation of the magnetic field,
the layer remains relatively thin even for adiabatic flows (see,
e.g., Baranov \& Zaitsev, 1995). The last isothermal case can be considered
as a limit of maximum conductivity, i.e. it mimics the effect of thermal condition.
Table\,\ref{tab:par} summarizes the parameters of our two model
calculations.

\begin{table}
\caption{Parameters of the model calculations.} \label{tab:par}
\begin{tabular}{ccccc}
\hline
No. & $\alpha$ & $M_{\rm ISM}$ & $M_{\rm A,ISM}$ & $\gamma$ \\
\hline
 1 & 90$\degr$ & 1.5 & 1.77 & 5/3 \\
 2 & 0$\degr$ & 1.89 & 1.77 & 1.0001 \\
\hline
\end{tabular}
\end{table}


\subsection{Plasma and magnetic field distribution}

\begin{figure*}
\includegraphics[scale=0.8]{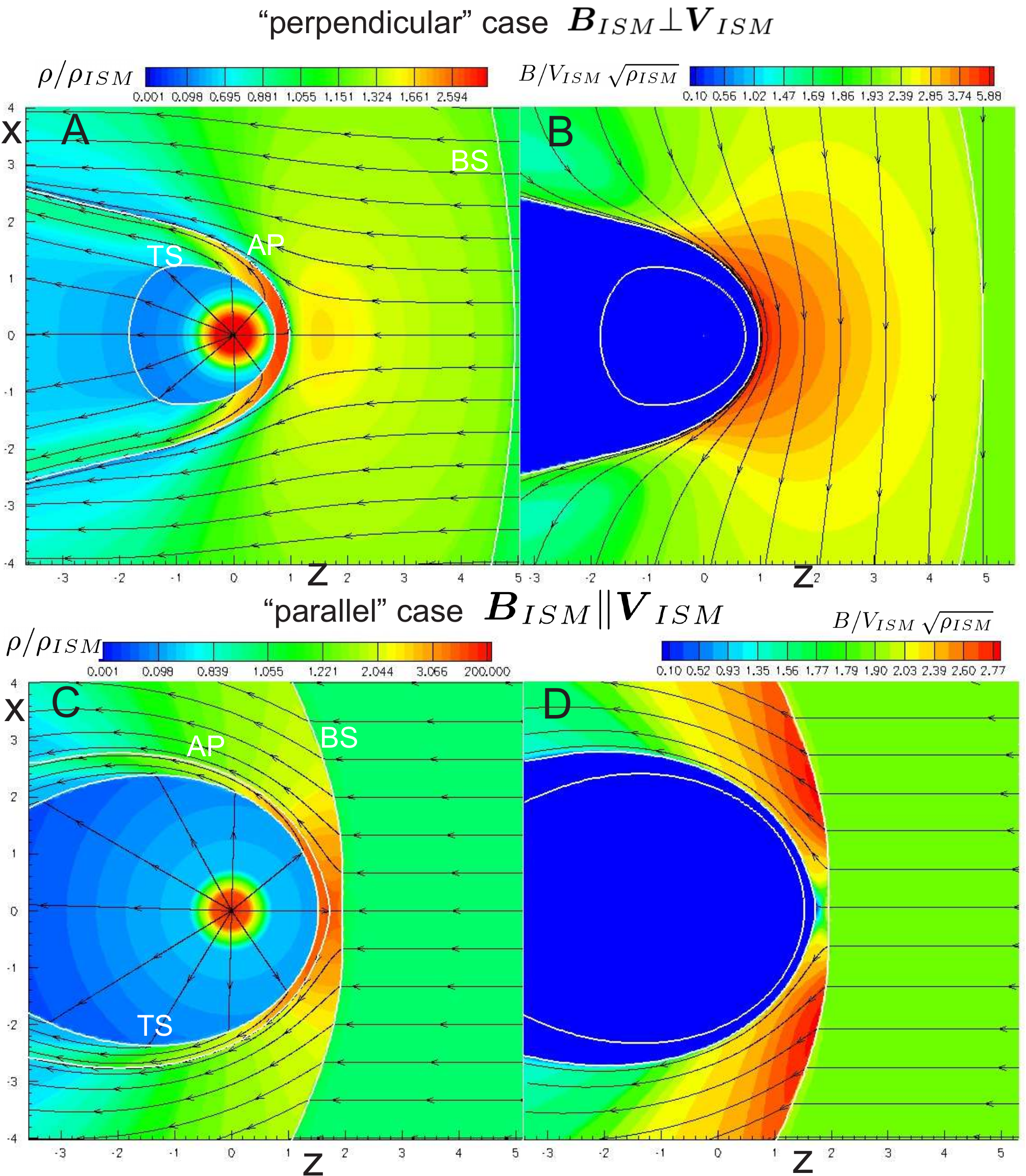}
\caption{2D distributions of the plasma density and streamlines
(panels A and C) and the magnetic field with the magnetic field
lines (panels B and D) in the $(ZX)$-plane for the perpendicular
(panels A and B) and parallel (panels C and D) cases. The termination
shock (TS), the astropause (AP) and the bow shock (BS) are plotted
with white lines. Distances along $X$ and $Z$ axes are normalized to $R_0$.} \label{plasma}
\end{figure*}

Fig.~\ref{plasma} presents the results of model calculations for
the two considered cases. The rectangular system of coordinates is
the following: the $Z$-axis is opposite to $\vecV_{\rm ISM}$ and
the $X$-axis lies in the plane $(\vecB_{\rm ISM}, \vecV_{\rm
ISM})$.
In the parallel case with $\vecB_{\rm ISM}
\parallel \vecV_{\rm ISM}$ the problem is axisymmetric. Detailed
analysis of the plasma distribution in the astrospheres
is beyond the scope of this paper and it will be performed
elsewhere. Here we briefly summarize the main features of the
flows.

Fig.~\ref{plasma} shows that both in the perpendicular and
parallel cases the plasma density is maximum between the TS and
the AP, and shows a secondary maximum at the AP from the
interstellar side. In the parallel case, the bow shock is
much closer to the star compared with the perpendicular one. This is
because of the assumed specific magnetic field configuration and
the nearly isothermal character of the flow in the parallel case.
In the perpendicular case, the magnetic field has a maximum at the
nose part of the astrosphere beyond the AP, while in the parallel
case there is a local minimum of magnetic field intensity at the
critical point. This is because in the parallel case $\textbf{v}_p
\sim \textbf{B}$ everywhere and $v_p=0$ at the critical point.
Distributions of the plasma velocity and magnetic field are critically important for
behavior of the dust grains with small gyroradii.

\subsection{Spatial dust distribution: formation of the filamentary structures}

\subsubsection{Perpendicular case}

At first, we calculate the dust distribution for the perpendicular
case. For the sake of simplicity, we do not consider a realistic
mass and size spectrum of dust grains in the ISM \citep[see, e.g.,
][]{mrn_1977}. Instead, we consider dust grains with several
different values of $a_d$ (see Table\,\ref{tab:ad}; parameters $r_{gyr}$ and $D_{gyr}$ presented in this Table as well are discussed in this section below).

\begin{table}
\caption{Parameters of the dust grains.} \label{tab:ad}
\renewcommand{\footnoterule}{}
 \begin{center}
 \begin{minipage}{\textwidth}
\begin{tabular}{ccc}
\hline
 $a_{\rm d}$ & $r_{\rm gyr}$ & $D_{\rm gyr}$ \\
\hline
 5.4 & 0.01 &  0.125 \\
 3.6 & 0.015  &  0.188 \\
 1.8 & 0.032  & 0.375 \\
 1.08 & 0.053 & 0.649 \\
 0.72 & 0.068 &  1.025 \\
 0.36 & 0.129 & 2.148 \\
 0.18 & 0.547 & 4.495 \\
 0.09 & 1.598 & 7.799 \\
\hline
\end{tabular}
\end{minipage}
\end{center}
 All parameters are dimensionless ($r_{gyr}$ and
$D_{gyr}$ are normalized to $R_0$). Gyroradius $r_{gyr}$ is
calculated at $z\approx1.6$ for the dust particle moving at
infinity along Z axis. $D_{gyr}$ is an averaged distance that a
dust grain passes along Z axis during one gyrorotational period.
\end{table}

Let us start from consideration of motion of the positively
charged dust grains under an influence of the electromagnetic
force $\textbf{F}_{\rm L}=q (\textbf{v}_d \times \textbf{B} +
\textbf{E})$, where $\textbf{E}=-\textbf{v}_p\times \textbf{B}$.
As mentioned before, outside of the BS the supersonic interstellar
plasma is undisturbed, i.e. both plasma and dust move together
with velocity $\vecV_{\rm ISM}$, so that $F_{\rm L}=0$. This means
that with approaching to the BS all dust grains move freely along
straight lines. Plasma flow decelerates at the BS, hence $v_d\neq
v_p$ after crossing the BS and the Lorentz force appears. The
Lorentz force initiates a gyrorotation of the dust grains around
magnetic field lines. Radius of the gyromotion (gyroradius or
Larmor radius) is $r_{gyr}=m\,v_{\perp}/|q|\,B$, where $v_{\perp}$
is the dust velocity component perpendicular to the magnetic field
vector. In dimensionless form
$\hat{r}_{gyr}=\hat{v}_{\perp}/(a_d\,\hat{B})$, meaning that
particles with the largest $a_d$ have the smallest gyrordius and
vice versa.

The motion of a dust particle can be represented as the gyromotion
around the magnetic field line frozen into the plasma and the
motion of the guiding centre. The guiding centre velocity
$\mathbf{V}_{gc}$ can be represented as a sum of the plasma
velocity $\mathbf{v}_{p}$ (which is the velocity of the
$\mathbf{E} \times \mathbf{B}$ drift) and the drift velocity
$\mathbf{V}^d$ caused by the spatial gradients of the magnetic and
electric fields, i.e. $\mathbf{V}_{gc} = \mathbf{v}_{p} +
\mathbf{V}^d$. Note that along the most part of the dust
trajectory the second drift velocity is negligible compared with
the first one for the most considered dust particles. That is, the velocity of the guiding centre is
almost equal to the plasma velocity.

\begin{figure*}
\includegraphics[scale=0.8]{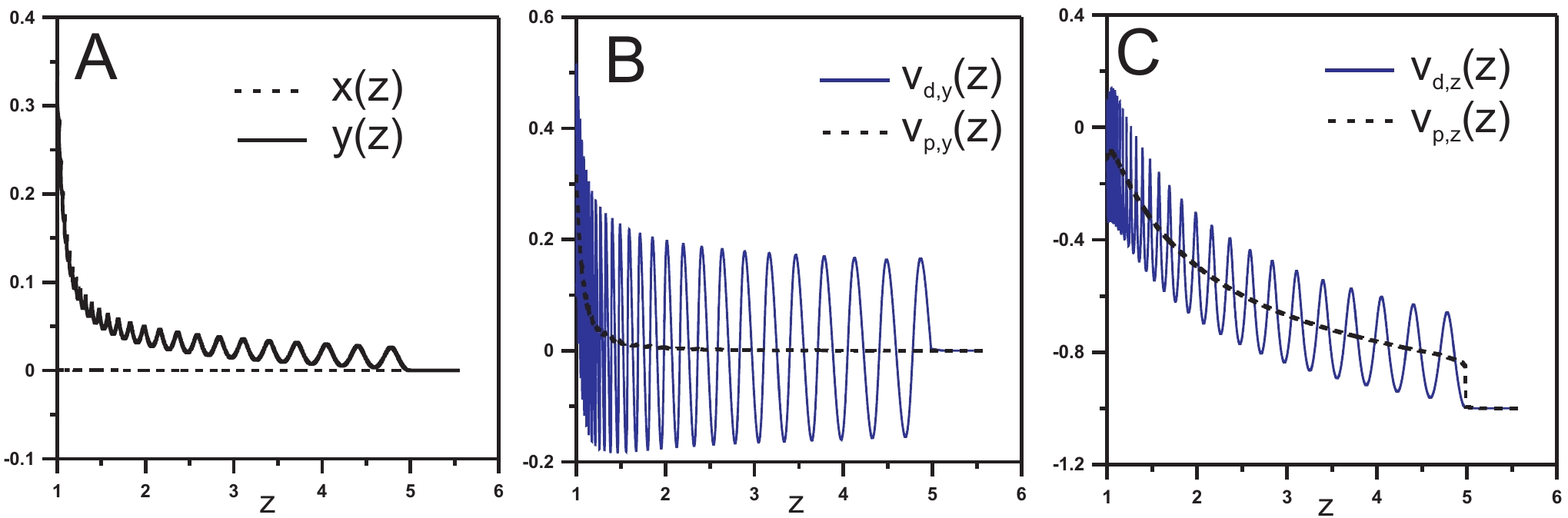}
\caption{Example of one trajectory of a dust grain with $a_d=5.4$.
The trajectory is started at $P_0$. Plot A shows $x(z)$ and
$y(z)$, plots B and C show $y$- and $z$-velocity components of the
plasma and dust along the trajectory. All distances are normalized to $R_0$, velocities are normalized to $V_{ISM}$.} \label{tra-upwind}
\end{figure*}

Fig.~\ref{tra-upwind}A shows a trajectory ($x(z)$ and $y(z)$) of a
dust grain (with the largest $a_d$ of 5.4) started at the point
$P_0$\,($x=0$, $y=0$, $z=5.5$). Hereafter, all presented
quantities ($x, \, y, \, z, \, n_d, \, v_p, \,  v_d$, etc) are
dimensionless: all distances are normalized to $R_0$,
all velocities are normalized to $V_{\rm ISM}$, and dust number density $n_d$ is normalized
to that in the ISM ($n_{d,ISM}$) for each magnitude of $a_d$. I.e. dimensionless number density of all particles
in the ISM is unity.
 We have omitted
the sign ``hat'' to simplify notations.
We consider only a part of the trajectory between the BS and the
AP, which are located, respectively, at $z\approx5$ and
$z\approx1$. Figs~\ref{tra-upwind}B and \ref{tra-upwind}C plot
the $y$- and $z$-components of the plasma ($\textbf{v}_p$) and
dust ($\textbf{v}_d$) velocities. The $x$-components of these
velocities are close to zero and we do not shown them. It is seen
that the dust particle rotates in the ($ZY$)-plane and is moving
together with the plasma
(i.e. the dust velocity oscillates around the plasma velocity).
$|v_{p,z}|$ decreases and $v_{p,y}$ increases with approaching to
the AP, which is caused by the plasma deceleration and flowing around the AP.

\begin{figure*}
\includegraphics[scale=0.8]{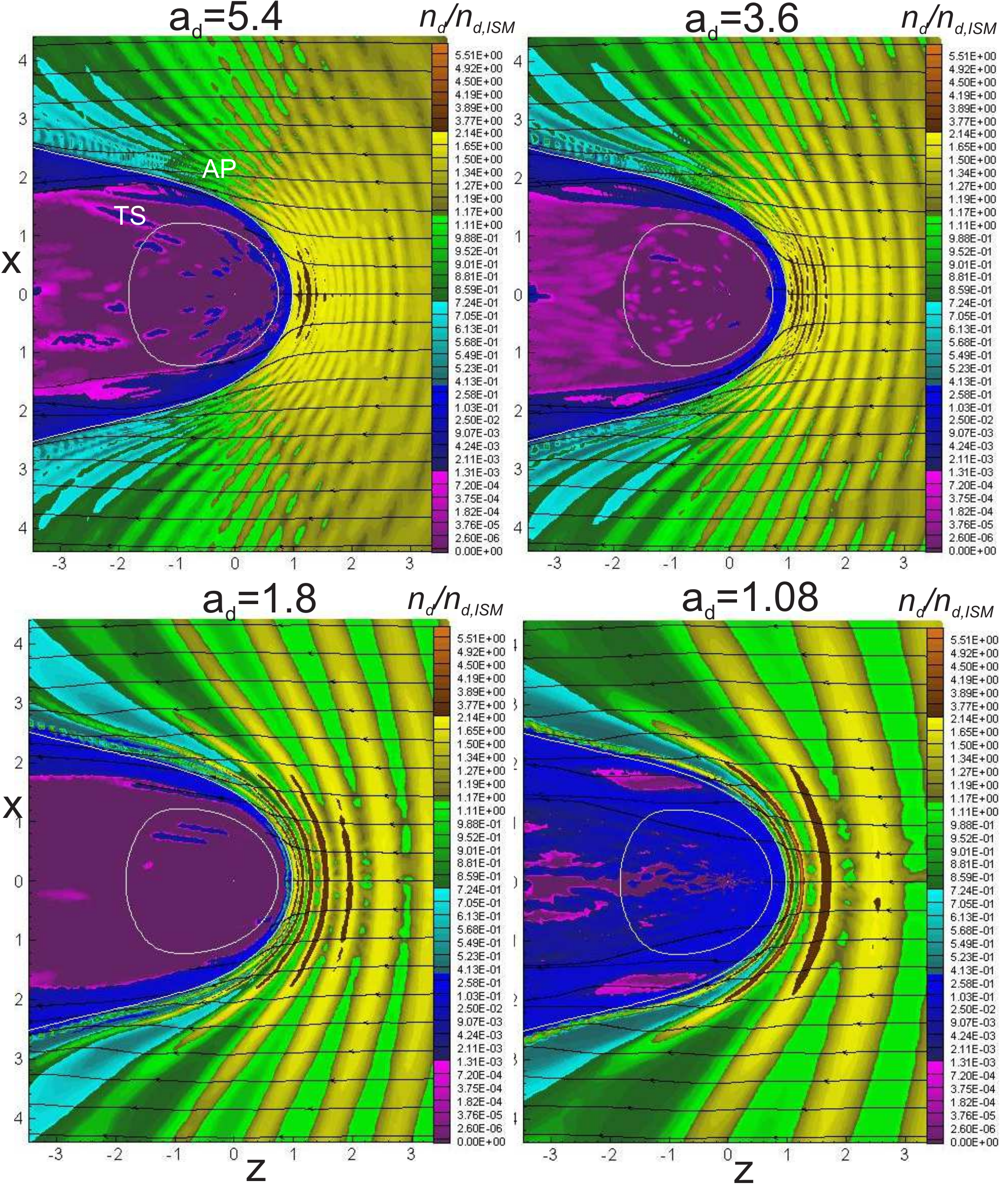}
\caption{2D distributions of the dust number density and
streamlines in the $(Z,X)$-plane for the dust grains with
different $a_d$ obtained for the perpendicular case. The termination
shock (TS) and the astropause (AP) are shown in all plots by white
lines. Distances are normalized to $R_0$.} \label{bv90-1}
\end{figure*}

\begin{figure*}
\includegraphics[scale=0.8]{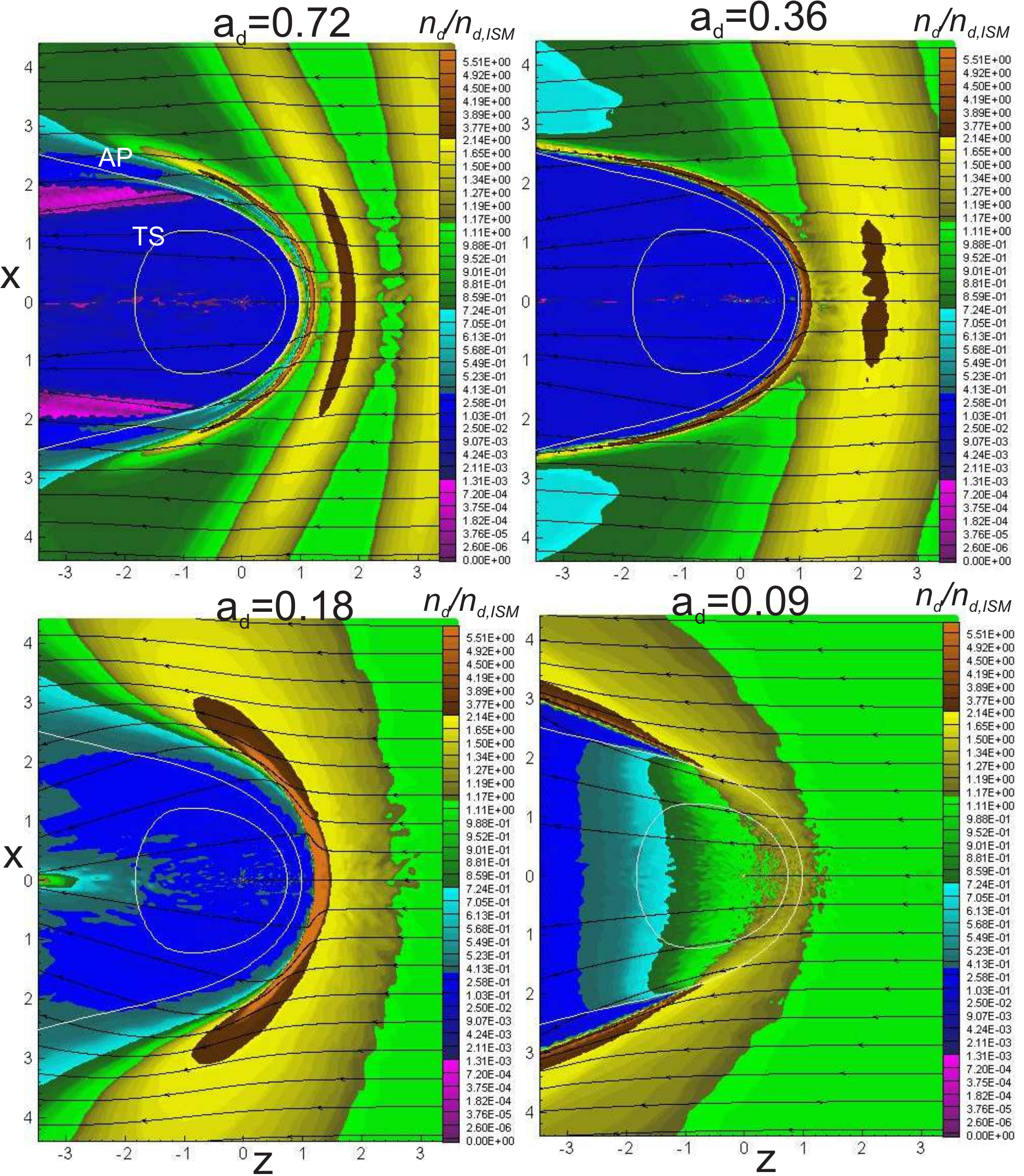}
\caption{2D distributions of the dust number density and
streamlines in the $(Z,X)$-plane for the dust grains with
different $a_d$ obtained for the perpendicular case. The termination
shock (TS) and the astropause (AP) are shown in all plots by white
lines. Distances are normalized to $R_0$.} \label{bv90-2}
\end{figure*}

Figs~\ref{bv90-1} and \ref{bv90-2} plot the dust number density ($n_d$)
and projection of the averaged dust velocity vector ($\textbf{V}_{d,av}$) in the $(ZX)$-plane.
One can see that the dust distribution in the astrosphere is
significantly non-uniform and non-monotonic. Several filamentary
or cirrus-like structures (alternating maxima and minima of the
dust number density) are clearly seen and their amount depends on
$a_d$. The ratio of the maximum number density to the minimum one
is about $2-4$.

\begin{figure*}
\includegraphics[scale=0.8]{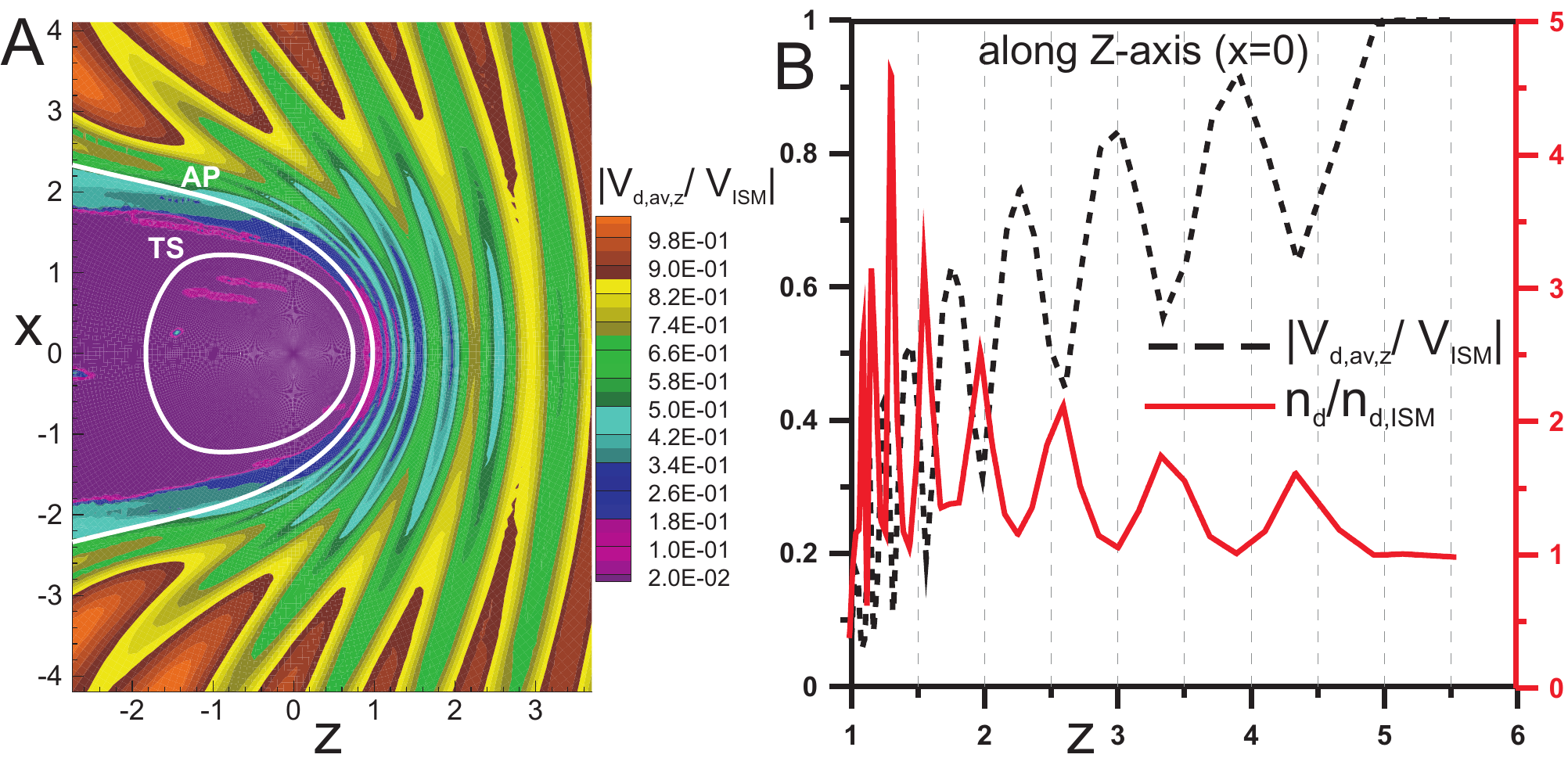}
\caption{ A. 2D distribution of the averaged dust z-velocity
component (absolute value) in the $(ZX)$-plane obtained for the
perpendicular case. B. Dust number density and averaged velocity
(absolute value of the $z$-component) along the $Z$-axis ($x$=0,
$y$=0).
All results are presented for the dust grains with
$a_d$=1.8. Distances are normalized to $R_0$.
} \label{bv90-1d}
\end{figure*}

\begin{figure*}
\includegraphics[scale=0.8]{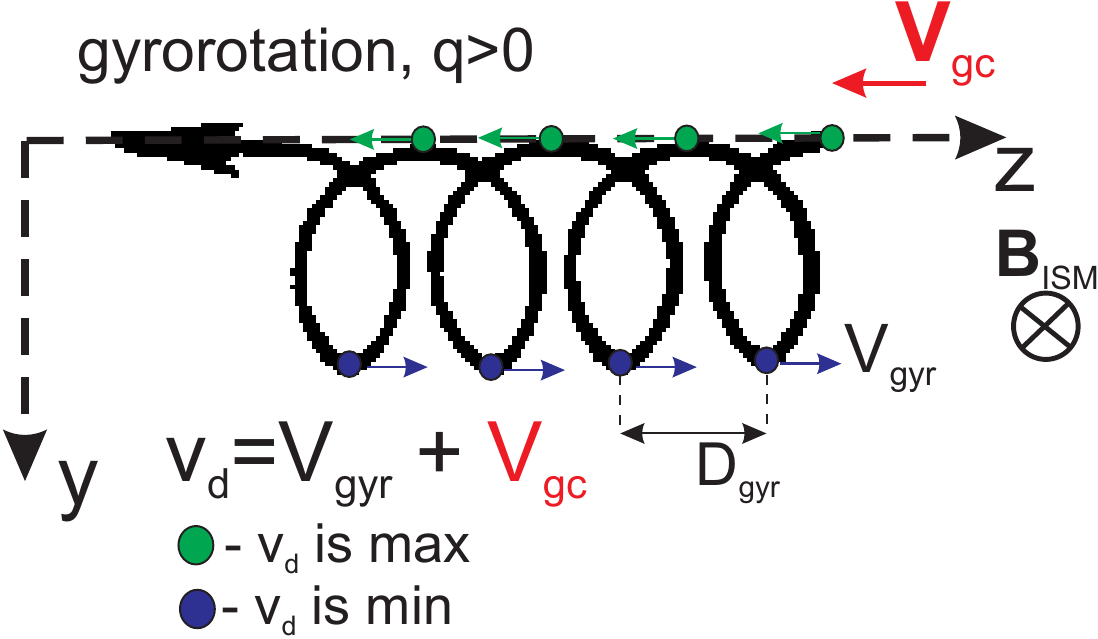}
\caption{Schematic picture of dust grain motion in the uniform
perpendicular magnetic field. Green dots correspond to maxima of
the local dust velocity, blue dots correspond to the local minima
(see text for details). } \label{gyromotion}
\end{figure*}

To understand the origin of these non-monotonic behavior of the dust number density let
us first consider the averaged velocity of dust grains with moderate
value of $a_d$=1.8: the 2D plot of $|V_{d,av,z}|$ in the
$(ZX)$-plane is shown in Fig.~\ref{bv90-1d}~A and the 1D plot in
the upwind direction is shown in Fig.~\ref{bv90-1d}~B (together with the dust number density).
From the 2D plot it is seen that the distribution of the dust
averaged velocity also has the filamentary structure with several minima
of the $|V_{d,av,z}|$-component. Positions of these velocity minima
correspond to positions of maxima of the number density
(it is clear from Fig.~\ref{bv90-1d}~B), i.e. dust accumulates in the regions of its deceleration. This is consistent with the
continuity equation. Thus, dust number density is not uniform in the astrosphere due to periodical deceleration and acceleration
of the dust grains.
This non-monotonic behavior of the dust velocity is caused by gyrorotation around magnetic field lines.
For example, for the trajectory started at point $P_0$ gyrorotation takes place in
 $(Z,Y)$-plane, and during each rotation at the point with maximal $y$,
 velocity vector of gyromotion is opposite to the plasma
velocity and, hence, the total individual dust velocity has a minimum;
oppositely, at the point with minimal $y$, total velocity has a maximum
 (see schematic illustration of this effect in
Fig.~\ref{gyromotion}).


The distance between the density maxima and their number depend
on the period of gyrorotation of dust grains (see
Fig.~\ref{gyromotion}). Namely, in the case of the uniform
electromagnetic field the distance between two adjacent maxima of
$y(z)$ along the trajectory is equal to $D_{\rm gyr}=v_{p,z}T_{\rm
gyr}$, where the dimensionless period of gyrorotation is $T_{\rm
gyr}=2\pi/(B a_{\rm d})$. In our case, when
$v_{p,z}$ and $B$ changes along the trajectory, a
distance between the density maxima depends on the averaged $D_{\rm gyr}$
 at each part of the trajectory (values of $D_{\rm gyr}$ averaged over the whole trajectory are given in Table~2).

\begin{figure*}
\includegraphics[scale=0.8]{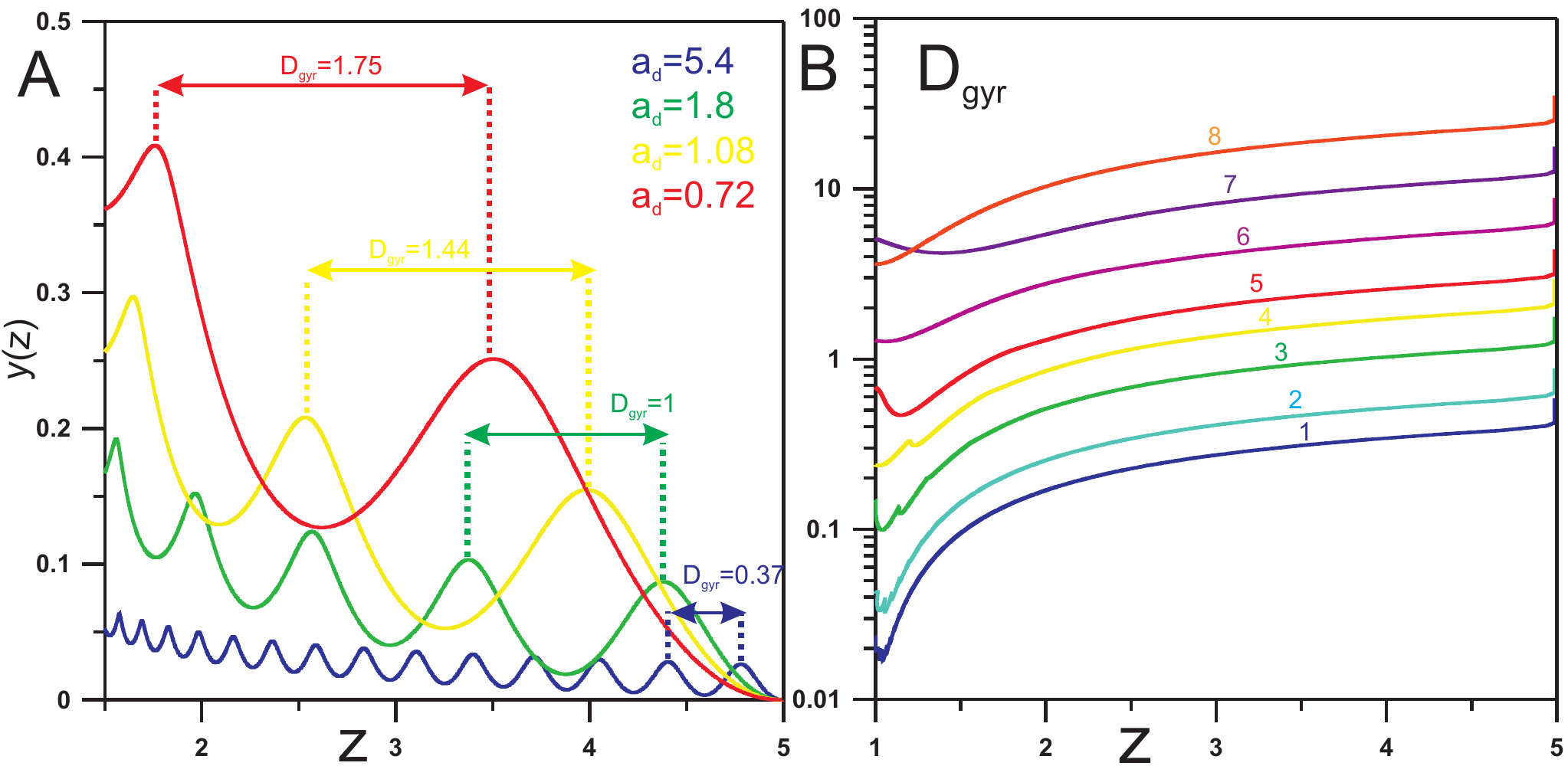}
\caption{A. The trajectory ($y(z)$) started at $P_0$ is shown for
dust grains with four different $a_d$. Periodical maxima and
minima of $y(z)$ show the gyromotion of the dust grains with
different gyroradii. The distance between adjacent maxima is
marked by $D_{gyr}$. B. Results of calculations of
$D_{gyr}=v_{p,z}\cdot T_{gyr}$ along the trajectory are shown for
all considered dust grains. All distances ($z,y,D_{gyr}$) are normalized to $R_0$.} \label{Dgyr}
\end{figure*}

Fig.~\ref{Dgyr}A shows $y(z)$ along the trajectories (started
at $P_0$) for different $a_d$. It is seen that the
grains are involved in the gyrorotation with corresponding
gyroradius (given in Table\,\ref{tab:ad}). Distance between
adjacent maxima of $y(z)$ along the trajectory
correspond to the distance between maxima of the dust number
density in Figs~\ref{bv90-1} and \ref{bv90-2}. Fig.~\ref{Dgyr}~B
shows $D_{gyr}$ as a function of $z$ for the trajectories started at $P_0$ and
the all considered values of $a_{\rm d}$. Note that $D_{\rm gyr}$
decreases with approaching to the AP because the magnetic field
$B$ increases (see Fig.~\ref{plasma}B). This is why the dust
density filaments become more frequent close to the AP (see
Fig.~\ref{bv90-1}).

Note that the dust grains with $a_{\rm d}>0.09$ almost do not
penetrate through the AP because of relatively small gyroradius.
The dust grains follow the magnetic field lines frozen in the
plasma and cannot cross the AP. Particles of fairly large size
($a_{\rm d}=0.09$, let us remind that $a_d$ is inversely proportional to square of grain's radius) have the dimensionless gyroradius of $\sim$1.6,
which this is enough to penetration inside the AP (see
Fig.~\ref{bv90-2}~D).

It is also interesting to examine Fig.~\ref{bv90-2}C (for $a_{\rm
d}=0.18$). It is seen that there is one pronounced maximum of the
dust number density and this maximum do not touch the AP.
Therefore, the following situation is possible. Let us imagine
that we observe an astrosphere due to emission of the interstellar
dust and see one filamentary structure corresponding to the
maximum of the dust number density. It is possible that the
position of this observed filament does not coincide with the AP
and the BS.
So, it is important for the data analysis that usually we cannot
observe the AP and the BS directly, but only what we see is the
dust distribution in the astrosphere, which can be quite
complicated. Detailed numerical MHD modelling of dust distribution
in astrospheres is, therefore, needed for correct interpretation
of observations.

\subsubsection{Parallel case}

Here we consider the parallel case, which is simpler than the
perpendicular one because the magnetic field vector is always
parallel to the plasma velocity vector \citep[see,
e.g.,][]{baranov_zaitsev_1995}. Therefore, $E=-\textbf{v}_p \times
\textbf{B} = 0$ and motion of the dust grains is a sum of
gyrorotation around magnetic field lines, motion along field lines
(with constant velocity $v_{\parallel}=\vecV_{\rm ISM}$) and
drifts due to spatial gradients of the magnetic field.

\begin{figure*}
\includegraphics[scale=0.8]{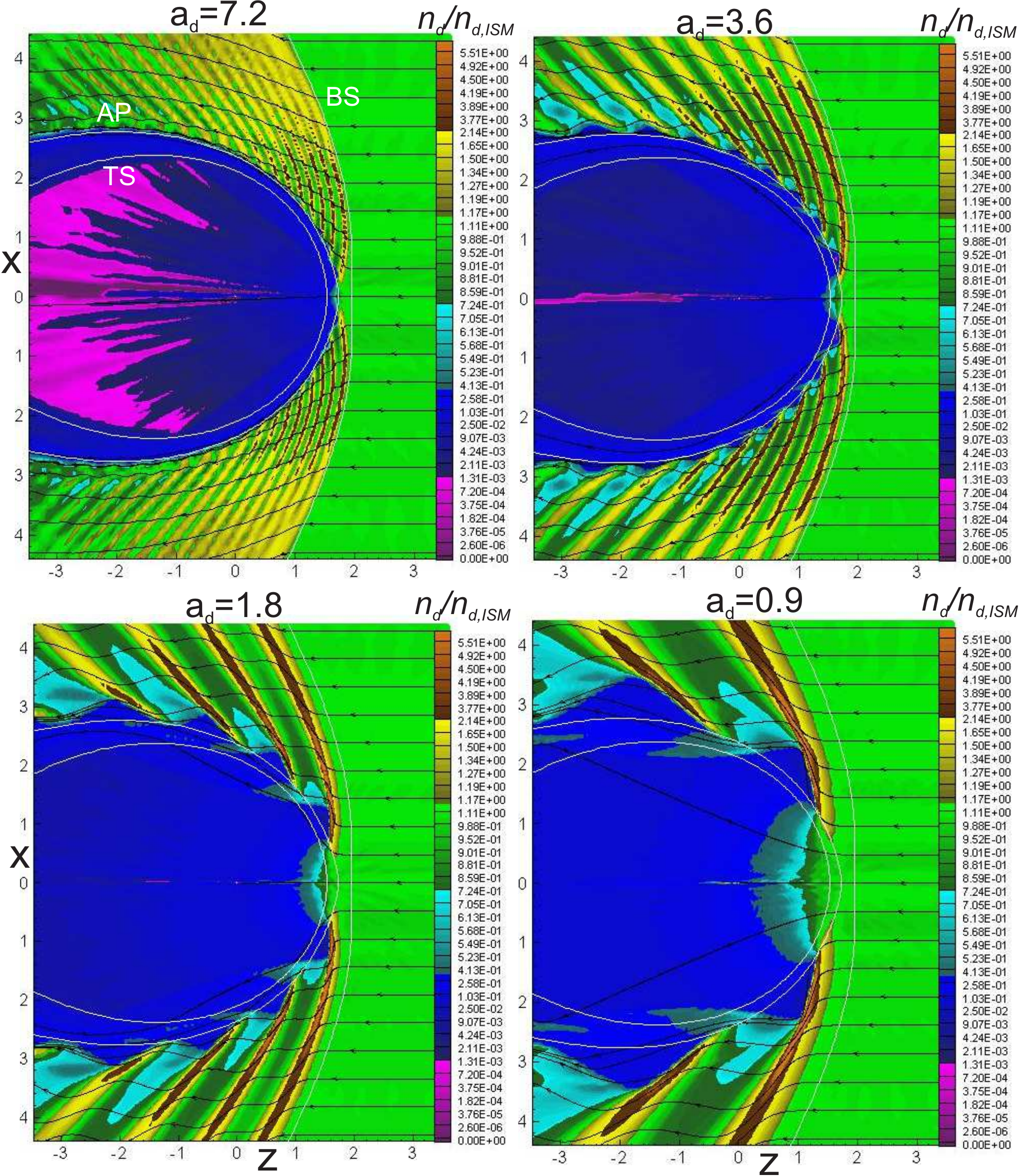}
\caption{2D distributions of the dust number density and
streamlines in the $(Z,X)$-plane for the dust grains with
different $a_d$ obtained in the parallel case. The termination shock
(TS), the astropause (AP) and the bow shock (BS) are shown in all plots by white
lines. Distances are normalized to $R_0$.} \label{bv0-2d}
\end{figure*}

Fig.~\ref{bv0-2d} shows the interstellar dust number density and
streamlines in the $(ZX)$-plane for $a_d=$7.2, 3.6, 1.8, 0.9. The
periodical filamentary structures are clearly seen in the
distributions. The physical mechanism of formation of such
structures is the same as discussed in the previous section and is
caused by the gyrorotation of the dust grains. It is interesting
that in the parallel case all dust grains can penetrate through
the nose region of the AP (i.e. the region around the critical
point) where the plasma velocity and the magnetic field are zero.

The presented results show that the filamentary dust distribution
in astrospheres can arise both in the perpendicular and parallel
cases. In the more general case of oblique magnetic field the
overall qualitative picture remains the same, but the emerging
filamentary structure is more complicated due to asymmetric
MHD-structure of the astrosphere.

\subsection{Shapes of astrosphere for different lines-of-sight}

The filamentary, cirrus-like structure of astrospheres was
revealed with modern infrared telescopes (see
Section\,\ref{sec:int}). Since the infrared emission of
astrospheres around hot stars is mostly due to the absorption of
stellar ultraviolet emission and its re-emission at longer
wavelengths (e.g. van Buren \& McCray 1988), it is naturally to
attribute the observed filamentary structures to the specific dust
distribution in astrospheres. The spectral characteristics of the
resulting infrared emission are determined by the dust grain's properties and
temperature, which, generally speaking, depends on the distance
from the central star. However, in this work we do not consider
the radiative transfer and its effect on the dust grains, which do
not allow us to determine the dust temperature and precludes us
from quantitative comparison of our model astrospheres with the
observed ones. To make things simpler, we assume that the dust
temperature is constant and does not depend on grain's size, so that the observed intensity of the
infrared emission is proportional to the column number density of
the dust particles, i.e. the integral of the dust number density
along our line-of-sight.

\begin{figure}
\includegraphics[width=8cm]{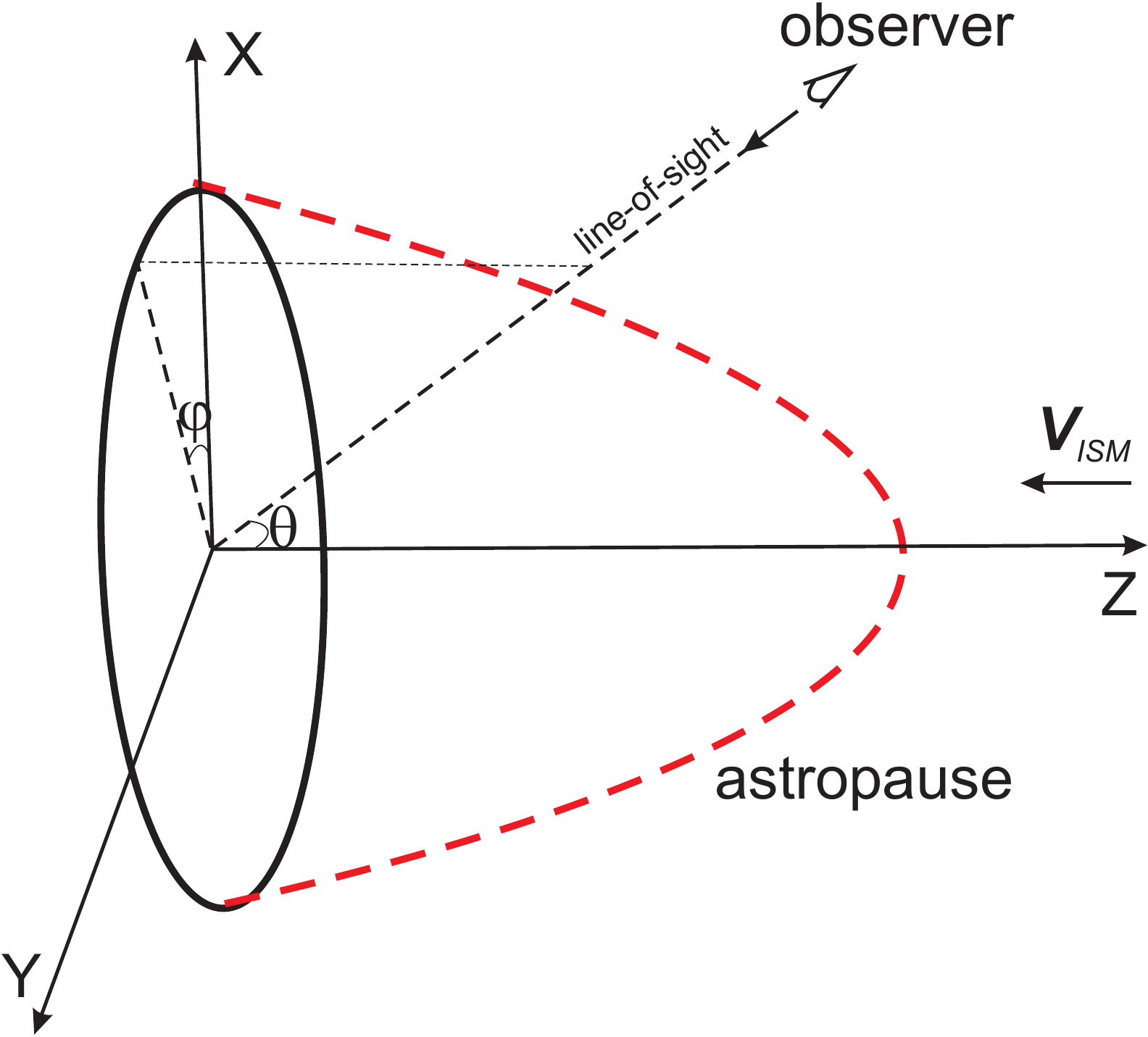}
\caption{Schematic representation of observation of an astrosphere
from an arbitrary line of sight. The astropause is shown by red dashed
line. Line-of-sight is characterized by two spherical angles:
$\theta$ and $\phi$.} \label{scheme}
\end{figure}

In previous sections we presented the spatial distribution of the dust
number density in the $(ZX)$-plane. Now we will consider how our
model astrospheres will appear for different lines of sight,
defined by angles $\theta$
and $\phi$
as specified in Fig.~\ref{scheme}.

\begin{figure*}
\includegraphics[scale=0.8]{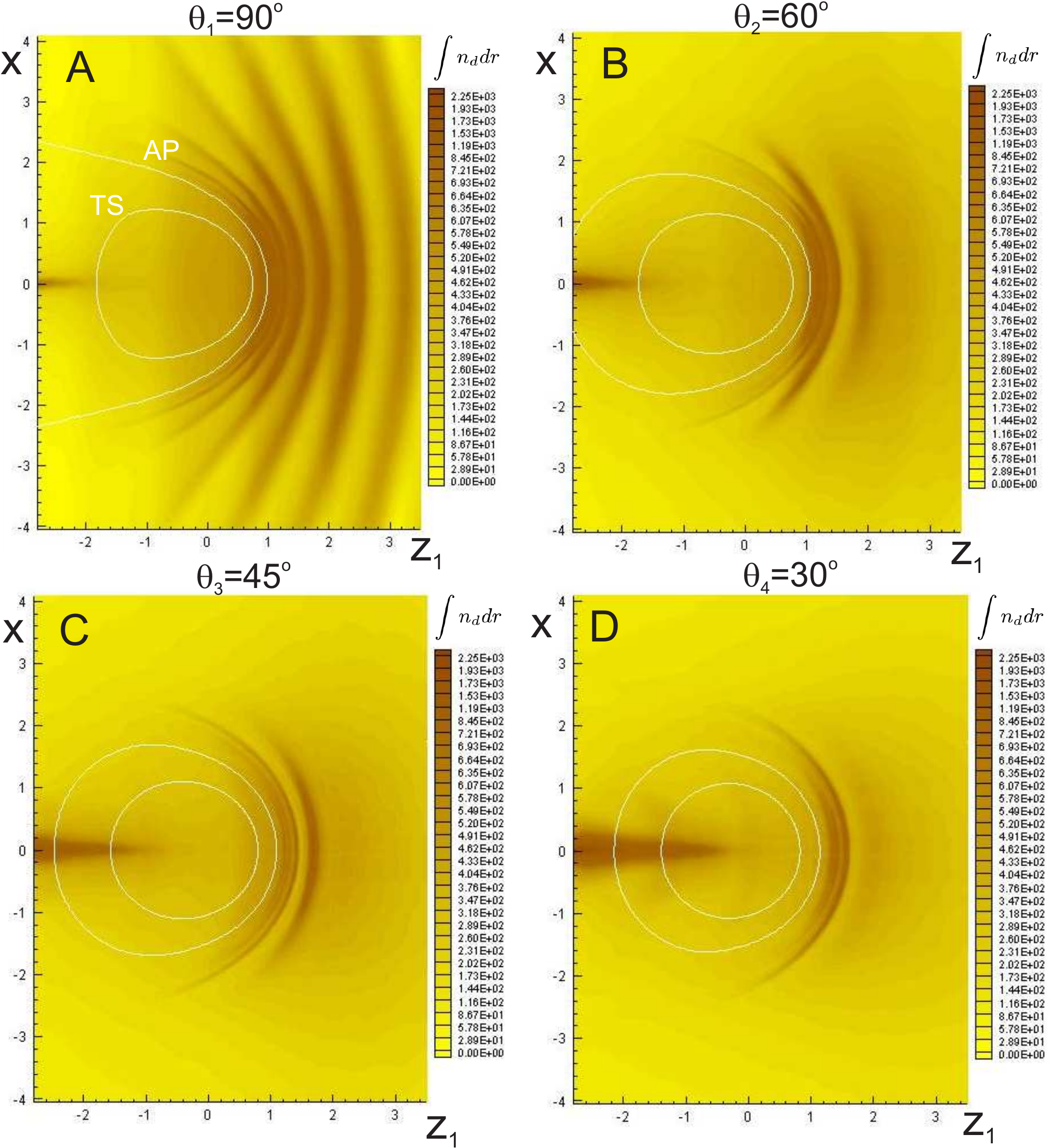}
\caption{2D maps of integrated dust number density along four
lines of sight in the $(ZY)$-plane ($\phi$=90$^{\circ}$), with
$\theta=90\degr$, 60$\degr$, 45$\degr$ and 30$\degr$.
Results are presented for the plane $XZ_1$ that is
perpendicular to the line-of-sight. Projections of the termination shock (TS) and astropause (AP)
on this plana are shown by white lines. The maps are
obtained for the perpendicular case and the dust grains with
$a_d=1.8$. Distances are normalized to $R_0$.
} \label{intensity}
\end{figure*}

For the sake of illustration, we choose four lines of sight in the
$(ZY)$-plane (i.e. $\phi$=90$^{\circ}$), with $\theta=90\degr$,
60$\degr$, 45$\degr$ and 30$\degr$.
Fig.\,\ref{intensity} plots 2D maps of integrated dust number
density along the chosen lines of sight. Calculations are
performed for the perpendicular case and the dust grains with
$a_d$=1.8. One can see that the number of observed filaments
depends on the orientation of the line of sight, i.e. for the larger
$\theta$ one can distinguish the larger number of filaments at
the nose part of the astrosphere.
This is caused by simple geometrical effect: when we observe an astrosphere with acute $\theta$ different filaments overlap
each other and are seen as one filament in the plane perpendicular to the line-of-sight.

 Note that enhancement of the intensity seen in the tail
part of the astrosphere is caused by the 3D features of the dust distribution.
Namely, the dust flows around the
astropause, accumulates at flanks in $(ZY)$-plane and produce such maxima in the observational plane due to an integration of the
number density along the line-of-sight.

The presented results show that the observed structure of
astrosphere significantly depends on the angle between the line of
sight and the direction of stellar motion. This means that for
correct interpretation of observations the knowledge of both the
3D distribution of dust grains in the astrosphere and the
velocity vector of the central star might be of crucial importance.

\subsection{Transformation of $a_d$ to the dust grain's radius}
\label{size}

\begin{table}
\caption{Radii of dust grains (in $\mu$m) as a function of $a_d$ for two sets of dimensional parameters
of the astrosphere. Set~1: $R_0$=0.1~pc, $n_{\rm ISM}$=0.1~cm$^{-3}$. Set~2: $R_0$=1~pc, $n_{\rm ISM}$=1~cm$^{-3}$.} \label{tab:size}
\begin{center}
\begin{tabular}{cccccccc}
\hline
$a_d$ & 5.4 & 3.6 & 1.8 & 1.08 & 0.72& 0.36& 0.18 \\
\hline
 \multicolumn{1}{c|}{set~1} & 0.13 & 0.16 & 0.23 & 0.30 & 0.37 & 0.53 & 0.74 \\
 \multicolumn{1}{c|}{set~2} & 0.76 & 0.94 & 1.33 &  1.71 & 2.10 & 2.97 & 4.20 \\
\hline
\end{tabular}
\end{center}
\end{table}

It is instructive to transform the dimensionless parameter $a_{\rm d}$ to the size of the dust grains in
order to check whether such grains may exist in the ISM. From equations~(\ref{R0}) and (\ref{ad}), one has:
\begin{equation}
  a_{d}=\frac{q}{m} \frac{ R_0 \sqrt{\rho_{\rm ISM}}}{c_0} = \frac{q}{mc_0}
  \sqrt{\frac{\dot{M} v_{\rm w}}{4\pi V_{\rm ISM}^2}}.
\end{equation}
Thus, to obtain the dimensional value of $q/m$, one needs to specify either the stand-off distance of the astrosphere
and the ISM plasma density, or the stellar wind mass loss rate and velocities of the stellar wind and the ISM flow.
In principle, this could be done for each particular astrosphere. Here we consider typical ranges of observed
stand-off distances $R_0\sim0.1-1$~pc and the ISM plasma number density $n_{\rm ISM}\sim0.1-1$~cm$^{-3}$.
The dust charge (in CGS units) can be expressed through the surface potential $U$ and the radius of a dust grain
$r_{\rm d}$, namely, $q=Ur_{\rm d}$. Assuming that the dust grains are spherical (with density $\rho_{\rm d}=$2.5~g\,cm$^{-3}$),
one has $q/m=3U/(4\pi\rho_{\rm d}r_{\rm d}^2)$. As mentioned before, in this work we assume a constant dust grain charge
(and surface potential) along the trajectory (this is discussed in the next section) and adopt the dust surface
potential of $U$=+0.75\,V, which is typical for the dust in the Local interstellar medium around the Sun
\citep{grun_svestka_1996}. The potential is usually positive due to an influence of accretion of protons,
photoelectric emission of stellar and interstellar radiation and secondary electron emission \citep[see, e.g.][]{kimura_mann_1998, akimkin_etal_2015}.

Table~\ref{tab:size} gives radii of dust grains for $a_d=0.18-5.4$ and two sets of values of parameters $R_0$ and $n_{\rm ISM}$
(set~1: $R_0$=0.1~pc, $n_{\rm ISM}$=0.1~cm$^{-3}$; set~2: $R_0$=1~pc, $n_{\rm ISM}$=1~cm$^{-3}$).
As it is seen from Figs \ref{bv90-1} and \ref{bv90-2}, the filamentary structure exists for these magnitudes of $a_d$ (for smaller and larger values of $a_d$
 the filaments either do not exist or non-visible).
For set~1 we obtain the grains radii of 0.13-0.74~$\mu$m, and for set~2 the grains radii are 0.76-4.2~$\mu$m.
In the classical MRN \citep{mrn_1977} size distribution of interstellar dust grains
the radii of the grains range from 0.005 to 1~$\mu$m for graphite and from 0.025 to 0.25~$\mu$m for silicate.
Therefore, the size ranges of the dust grains required for the filamentary structure in our model is within the MRN size-distribution for set~1 and has intersection with the MRN distribution for set~2.
Also, Wang et al. (2015) stated that there are several independent observational evidences of presence of the very large dust grains in the ISM. They constrained the size distribution of the $\mu$m-sized dust population by fitting the observed mid-IR extinction at $\sim$3-8 $\mu$m in terms of the silicate-graphite-PAH model together with an extra population of $\mu$m-sized grains. They obtained the size-distribution of dust at the following ranges: 0.0001-0.4 $\mu$m for silicate and PAH, and two ranges 0.0001-0.4 $\mu$m and 0.5-6 $\mu$m for graphite.
Therefore, the size ranges of the dust grains required in our model for formation of filamentary structures seems to be
realistic especially in view of the facts that small dust grains can be swept out from astrospheres (Ochsendorf et al. 2014)
or destroyed (Pavlyuchenkov et al. 2013) by the stellar radiation.

\section{Discussion: limitations of the model}
\label{sec:discussion}

Now we discuss several limitations of our model, which we are going to improve in the future.

The first one is the neglecting of the drag force due to interaction of dust grains with protons
and electrons through direct and Coulomb collisions \citep{draine_salpeter_1979}.
Several authors take this force into account in simulations of the interstellar dust motion
in astrospheres and H\,{\sc ii} regions \citep[see, e.g.,][]{ochsendorf_etal_2014, akimkin_etal_2015}.
The drag force depends on the dust radius, relative velocity between plasma and dust ($v_{rel}$), plasma number density ($n_p=\rho/m_p$) and temperature (T).
Namely, $F_{drag}=r_d^2 \, n_p \, k T \cdot \hat{G}(v_{rel}, T)$, where $k$ is the Boltzmann constant, $\hat{G}$ is dimensionless function, presented e.g. by \citet{draine_salpeter_1979}. With this force the dimensionless motion equation~(\ref{motion_br}) can be rewritten as:
\begin{equation*}
 \dot{\hat{\textbf{v}}}_d = a_{d}
[(\hat{\textbf{v}}_d-\hat{\textbf{v}}_p)\times \hat{\textbf{B}}] + b_d \cdot \hat{p} \, \hat{G}(\hat{v}_{rel}, \hat{T}),
\end{equation*}
where dimensionless coefficient related to the drag force is the following:
\begin{equation*}
 b_d=\frac{R_0 \cdot \rho_{ISM} \cdot r_d^2}{m}.
\end{equation*}
Therefore, with substitution of equation~(\ref{ad}) and expression for $q=U\cdot r_d$ one has:
\begin{equation}
 \frac{b_d}{a_d}=\frac{R_0 \, \rho_{ISM} \, r_d^2 \, c_0}{U \, r_d \, \sqrt{\rho_{ISM}} \, R_0} = \frac{c_0}{U} \cdot \sqrt{\rho_{ISM}} \, r_d.
\end{equation}
Thus, an importance of the drag force relative to the Lorentz force is larger for bigger grain radii and plasma density.

\begin{figure}
\includegraphics[width=8cm]{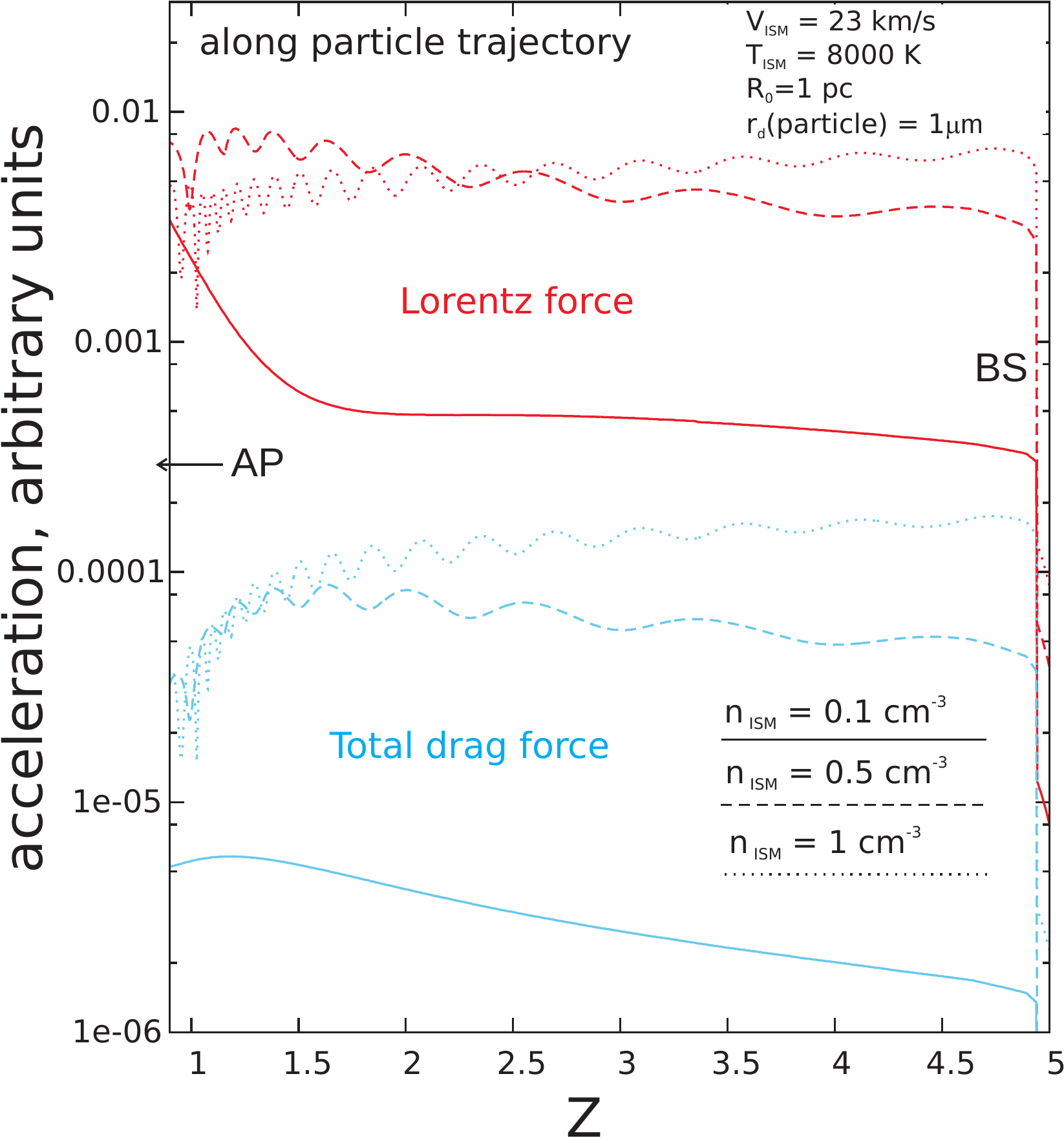}
\caption{Acceleration of a dust grain due to Lorentz and drag forces along its trajectory. Radius of grain is 1~$\mu$m.
Results are presented for three magnitudes of the ISM plasma number density
(solid curves correspond to $n_{ISM}$=0.1~$cm^{-3}$, dashed curves correspond to $n_{ISM}$=0.5~$cm^{-3}$,
and dotted curves correspond to $n_{ISM}$=1~$cm^{-3}$).}  \label{acceleration}
\end{figure}

We estimated acceleration of a dust grain (of a quite large radius of $r_d$=1~$\mu$m) due to the drag and Lorentz forces
separately for the typical ISM parameters: $T_{\rm ISM}=8\,000$~K, $n_{\rm ISM}=0.1-1$~cm$^{-3}$,
$V_{\rm ISM}$=25~km\,s$^{-1}$ and $R_0$=1~pc. The results are
shown in Fig.~\ref{acceleration}. It is seen that even for $n_{\rm ISM}=1$~cm$^{-3}$ the acceleration related with
the drag force is almost two orders of magnitude smaller than the acceleration caused by the Lorentz force.
Thus, the drag force can only be important either for very large grains (with radii of tens of microns) or for
very dense ISM (with $n_{\rm ISM}\gtrsim10$~cm$^{-3}$). Since none of these cases is considered in our
simulations, we neglect the drag force.

The second limitation of our model is the assumption of constant dust grain charge along the trajectory.
The charge of dust grains is controlled by three main processes: impinging of plasma particles (protons and
electrons), secondary electron emission due to electron impacts (this process is important in regions of hot plasma
with $T\ga10^5$\,K) and photoelectron emission due to stellar and interstellar radiation.
In principle, our model allows us to take changes of charge into account \citep[we did this in the case of the heliosphere, see][]{alexash_etal_2016}. Our study of dust charge in the heliosphere shows that the largest changes occur in the solar wind region
(i.e. inside the heliopause, which is an analogue of the astropause, see Fig.~2 in Alexashov et al., 2016). In this work we do not
consider the dust dynamics inside the astropause. We estimate variations of the charge due to changes of the plasma density and temperature
at the bow shock and in the interaction region
between the BS and the AP, and find that they are not larger than 30~\%. For simplicity, we neglect these changes
and assume that the dust charge is constant. Note that in the case of massive hot stars the dust charge can be mostly influenced by the stellar radiation,
which decreases with distance from the star as $1/r^2$. This may lead to changes of charge in the astrosphere if the layer between the BS and the AP is thick enough.
  Corresponding estimations of this effect should be done for each certain astrosphere. It should be noted also that possible variations of the charge will change
the gyroradius of dust grains, which in turn will change the distance between the filaments.

The third aspect that should be considered further is the realistic size distribution for the ISM dust grains.
In this work we perform calculations for dust grains with a fixed size (characterized by the chosen parameter $a_{\rm d}$) separately, while in
reality there are dust grains with a range of sizes \citep[see, e.g.,][]{mrn_1977, wang_etal_2015}. The standard MRN size distribution
is a $r_d^{-3.5}$ power law in grain radii, i.e. amount of the smallest grains is the largest.
Thus, if we will calculate the dust column number density from particles of all sizes, then the contribution from the smallest
grains will prevail and no filaments will appear (because the filaments produced by these grains are very narrow and the separation
between them is too small). However, the effect of the small grains on the appearance of filaments would be negligible
if they are swept out from astrospheres \citep{ochsendorf_etal_2014} or destroyed \citep{pavlyuchenkov_etal_2013} by the stellar radiation.

And the last limitation is that the observed intensity of the infrared emission from astrospheres depends not only on the dust column number density, but also on the dust temperature, which in its turn depends on the distance from the star and the physical parameters (size, composition) of the dust grains
(see, e.g., Fig.~3 in Wang et al., 2015 and Eq.~(14) in Decin et al., 2006). Hence, the relative contribution of the small and big dust grains to the total emissivity may be different and not simply proportional to their number densities.
A radiative transfer model is needed for proper modelling of the dust heating and radiation \citep{pavlyuchenkov_etal_2013, mackey_etal_2016}.
We are going to incorporate the radiative transfer code to our model in future works.

%

\section{Summary}
\label{sec:sum}

In this work we investigate the spatial interstellar dust distribution in
astrospheres with the interstellar magnetic field
perpendicular and parallel to the velocity vector of the
interstellar flow. We suggest a new physical mechanism for
formation of filamentary structures observed for some
astrospheres. It is shown that the alternating minima and maxima of
the dust density occur between the astrospheric bow shock and the
astropause due to periodical gyromotion of the dust grains. These
filamentary structures appear in the model for particles with the period
of gyration comparable to the characteristic time of the
dust motion between the bow shock and the astropause.

More realistic numerical modelling and comparison with the observational data will be done in the future works.



\section*{Acknowledgments}
This work is supported by the Russian Science
Foundation grant No.~14-12-01096.

%
%
%
%

\clearpage

\end{document}